\documentclass[12pt]{article}
\usepackage{latexsym}
\usepackage{hyperref}

\def\NON{\nonumber\\}
\def\NXT{\\}

\def\a{\alpha}

\def\c{\chi}
\def\d{\delta}
\def\g{\gamma}

\def\j{\psi}

\def\m{\mu}
\def\n{\nu}

\def\t{\tau}

\def\L{\Lambda}

\def\P{\Pi}

\def\ca{{\cal A}}
\def\cb{{\cal B}}

\def\cd{{\cal D}}

\def\cm{{\cal M}}

\def\cbo{{\,\raise-.15ex\Sc [\,}}                       

\def\sl#1{\rlap{\hbox{$\mskip 1 mu /$}}#1}      
\def\Sl#1{\rlap{\hbox{$\mskip 3 mu /$}}#1}      

\def\svev#1{\left\langle #1\right\rangle}       

\def\ddt#1{{\buildrel {\hbox{\LARGE .\kern-2pt.}} \over {#1}}}

\def\sstyle{\scriptstyle}

\def\ie{\mbox{\it i.e.} }
\def\eg{\mbox{\it e.g.} }

\def\frac#1#2{ {\sstyle {#1\over #2} } }

\thicklines                         
\def\bj{\overline\psi}
\def\bc{\overline\chi}

\def\bq{\overline{q}}

\def\tx{\tilde{x}}

\def\Dst{D_{s}}
\def\Drg{D_{\rm rg}}

\def\textit#1{{\it \!\!\! #1 \!\!}}

\def\be{\begin{equation}}
\def\ee{\end{equation}}

\def\sssec#1#2{
  \vskip 1ex
  \noindent {\bf #1.} {\it #2.}}

\begin{document}
\hyphenation{fer-mio-nic per-tur-ba-tive pa-ra-me-tri-za-tion
pa-ra-me-tri-zed a-nom-al-ous}

\renewcommand{\thefootnote}{*}

\begin{center}
\vspace{10mm}
{\large\bf
Renormalization-group blocking the fourth root
\\[2mm]
of the staggered determinant\footnote{
  Talk presented at the
  23rd International Symposium on Lattice Field Theory,  25-30 July 2005,
  Trinity College, Dublin, Ireland, and at the
  Workshop on Computational Hadron Physics,
  Nicosia, Cyprus, September 14-17, 2005.}
}
\\[12mm]
Yigal Shamir
\\[8mm]
{\small\it School of Physics and Astronomy\\
Raymond and Beverly Sackler Faculty of Exact Sciences\\
Tel-Aviv University, Ramat~Aviv,~69978~ISRAEL}\\
{\tt shamir@post.tau.ac.il}
\\[10mm]
{ABSTRACT}
\\[2mm]
\end{center}

\begin{quotation}
Lattice QCD simulations with staggered fermions rely
on the ``fourth-root trick.'' The validity of this trick
has been proved for free staggered fermions using renormalization-group
block transformations. I review the elements of the construction
and discuss how it might be generalized to the interacting case.
\end{quotation}

\renewcommand{\thefootnote}{\arabic{footnote}}
\setcounter{footnote}{0}

\newpage
\sssec{1}{The fourth-root trick}
A staggered-fermion field $\c(x)$
consists of a single fermionic degree of freedom per color per lattice
site \cite{ks}. The sixteen variables (per color)
residing in each $2^4$ hypercube of the euclidean lattice
can be re-grouped into four quark fields in the continuum limit.
Thus, a single staggered field could be used to simulate the up, down,
strange and charm quarks, all at once, provided an appropriate mass matrix
is chosen \cite{mgjs}. For various technical reasons, however, this is not
the way staggered-fermion simulations are done.

The ``rooted'' Boltzmann weight used to generate the dynamical
staggered-fermion configurations contains a factor of ${\rm det}^{1/4}(\Dst+m)$
for each of the three light quarks, where $\Dst$ is the massless,
interacting, anti-hermitian, one-component,
staggered-fermion operator.\footnote{
  In practice the one usually sets $m_u=m_d=m_{\rm light}$ and simulates
  ${\rm det}^{1/2}(\Dst+m_{\rm light})\,
   {\rm det}^{1/4}(\Dst+m_{\rm strange})$.
}
The {\it raison d'etre} behind the fourth-root trick is simple \cite{trick}.
In the continuum limit,
this staggered-fermion determinant will describe four equal-mass quarks
that interact only through the exchange of gluons; hence, the determinant's
fourth root is expected to account for a single quark with the same mass.
The difficultly is that, for any finite lattice spacing $a$, the four quark
species (or four ``tastes,'' reserving the term ``flavors'' to
the different-mass species) are entangled at short distances.
There is no simple way to represent the rooted Boltzmann weight
as a path integral with a local action. This raises the question whether
the rooted staggered theory is consistent with the rules of local
quantum field theory, and whether its continuum limit is in the same
universality class as QCD. The unprecedented accuracy of staggered-fermions
simulations has made this issue all the more imminent \cite{milc}.

In the free theory,
the re-grouping alluded to in the first paragraph is explicitly given by
\cite{taste}
\begin{equation}
  \j_{\a i}(\tx)
  = \sum_{r_\m=0,1}
  (\g_1^{r_1} \g_2^{r_2} \g_3^{r_3} \g_4^{r_4})_{\a i}\,
  \c(2\tx+r)\,.
\label{chitoq}
\end{equation}
Here $\a$ and $i$ are respectively Dirac and taste
indices, that both run from 1 to 4. The taste-representation variables
$\j_{\a i}$ live on a coarse lattice, with spacing $2a$,
whose sites are labeled
by the coordinates $\tx$. Each site of the original, fine lattice
has a unique representation
as $x=2\tx+r$, where $r_\m=0,1$. In the taste basis
the free (massless) staggered action is $S_0 = \bj D_0 \j$,
where
\begin{equation}
  D_0 = a^{-1} \sum_\m \Big([\g_\m \otimes I] i\sin(p_\m a)
  + [\g_5 \otimes \t_5 \t_\m] (1-\cos(p_\m a) \Big) \,,
\label{Dfree}
\end{equation}
\vskip -1ex \noindent
in momentum space.
The two sets of Dirac matrices, $\g_\m$ and $\t_\m=\g_\m^*=\g_\m^T$, act
on the Dirac and taste indices respectively.
$I$ is the identity matrix in taste space.
The kinetic term---the first term on the right-hand side---is
the naive discretization of the continuum $\Sl{D}$.
The second term lifts the doublers at the $p \ne 0$ corners
of the Brillouin zone (of the coarse lattice).
We will refer to it as a skewed---because of
the extra $\t_\m$ matrices---Wilson term.

As a start, there are two complementary ways to convince oneself that the
fourth root ``has to'' work in the free theory. First, one can look for
the four-fold degeneracy expected from
the taste structure. In momentum space,
the continuum limit of a free theory corresponds
to $|p|a \to 0$. The kinetic term starts off as $\sl{p}+O(a^2p^3)$,
whereas the skewed Wilson term starts off as $O(ap^2)$;
thus the skewed Wilson term drops out, and the expected four-fold degeneracy
is obtained.
Equivalently, one can examine the behavior of the propagator
$G_0=D_0^{-1}$ at large space-time separations.
The momentum-space representation of $G_0$ is easily worked out.
The only singularity, occurring at $p\to 0$, has the same structure
as in the continuum. The taste-breaking terms, originating in
the skewed Wilson term, do not produce a singularity in the propagator.
As a result \cite{cmp}, the taste-breaking part of the
coordinate-space free propagator vanishes exponentially with the separation,
with an $O(1)$ decay rate in lattice units.
We must keep in mind, though, that for momenta or separations
which are $O(1)$ in lattice units, the skewed Wilson term
does spoil the diagonal taste structure. Therefore we should not expect
that merely taking the fourth root of the product of all the eigenvalues,
small and large alike, will yield a sensible lattice operator.

Next let us examine one-loop perturbation theory, taking the vacuum
polarization $\P_{\m\n}(p)$ as an example.
We recall that the interacting theory is defined using the one-component
formulation, and that all of its symmetries are needed to
ensure that mass terms will always renormalize multiplicatively \cite{mgjs}.
How to construct an interacting theory in the taste basis is an issue
that we discuss later. For massless staggered fermions
one obtains
\begin{equation}
  \P_{\m\n}(p) = (\d_{\m\n} p^2 - p_\m p_\n)
  (4 c \log(pa) + \mbox{contact terms})\,.
\label{vacpol}
\end{equation}
The logarithmic term is universal. The coefficient $c$ is
the same as one would obtain for a single quark in a continuum calculation,
or for that matter, in a lattice calculation
using any one-flavor Dirac operator, such as a Wilson fermion.
The non-analytic, {\it long-distance} part of the vacuum polarization
of staggered fermions will therefore be reproduced by (say)
four Wilson fermions.
Taking the fourth root would replace these four Wilson fermions by one,
implying that the fourth root of the long-distance
part of the determinant is a sensible quantity.

Like in any perturbative lattice calculation, one-loop diagrams
with staggered fermions inside the loop also contain contact terms,
which are all local and gauge invariant.
The first of those will correspond
to (a lattice discretization of) $F_{\m\n}^2$, and has the effect of
a finite renormalization of the coupling constant. Other terms correspond
to irrelevant operators.
Similar statements apply to a calculation with Wilson fermions.
However, the contact terms are not universal,
and the staggered-fermions contact terms are {\it not} equal to four
times the Wilson-fermions contact terms. But then, this {\it short-distance}
discrepancy is not a problem: In lattice perturbation theory,
the fourth root of the staggered determinant will be almost equal to a
Wilson-fermion determinant; and, if we wish to, we can always make up for the
difference in the contact terms by introducing innocuous, local modifications
to the gauge-field action.

The fourth-root trick is also valid within the framework of low-energy
effective lagrangians that capture the infra-red limit of QCD \cite{PQ}.
The remaining challenge is to promote the arguments
to a fully non-perturbative setup. The lesson is that we will need
a device that can separate out long-distance from short-distance physics,
thus allowing us to treat differently the two parts of the  determinant.
This is precisely what Renormalization-Group (RG)
blocking was designed for \cite{bw,ph}.

\sssec{2}{Renormalization-group transformations}
Our first task is to establish the validity of of the fourth-root
trick in the free theory, using RG block transformations \cite{rg}.
A single RG blocking step works as follows \cite{bw}:
\begin{eqnarray}
  Z &=& \int d\j d\bj\; \exp(-\bj D_0 \j)
\NON
  &=& \int d\j d\bj dq d\bq\; \exp\left(
  -\bj D_0 \j
  - \a(\bq - \bj Q^\dagger)(q - Q \j) \right)
\NON
  &=& {\rm det}(G_1^{-1}) \int dq d\bq\; \exp(-\bq D_1 q) \,.
\label{RG}
\end{eqnarray}
The blocking parameter $\a$ has mass dimension one.
Here $\j,\bj$ live on the original lattice while $q,\bq$ live on the blocked
lattice. On the second line, the integrand has been multiplied by one;
on the last line, the $\j,\bj$-integration has been carried out.
The induced operator on the coarse lattice is $D_1$, and
$G_1^{-1} = D_0 + \a Q^\dagger Q$. We will assume that the blocking
step replaces each $2^4$ hypercube by one site of the coarse lattice.
The blocking kernel corresponds to the arithmetic
mean of $\j$ over a $2^4$ hypercube,
\begin{equation}
  (Q\j)(\tx) = 2^{-4} \sum_{r_\m=0,1} \j(2 \tx_\m + r_\m)\,.
\label{Q}
\end{equation}
If we apply the blocking transformation $n$ times,
the result is similar to  Eq.~(\ref{RG}) with $G_1 \to G_n$,
$D_1 \to D_n$, where (see ref.~\cite{rg} for further explanations)
\begin{equation}
  G_n^{-1} = D_0 + \a_n Q_n^\dagger Q_n  \,, \qquad
  D_n^{-1} = \a_n^{-1} + Q_n D_0^{-1} Q_n^\dagger \; .
\label{DGn}
\end{equation}
Here $Q_n=Q^{(n)} Q^{(n-1)}\cdots Q^{(1)}$
and $1/\a_n = 1/\a^{(n)} + 2^{-4}/\a^{(n-1)} + 2^{-8}/\a^{(n-2)} + \cdots$,
where $Q^{(j)}$ and $\a^{(j)}$ are respectively the blocking
kernel and parameter used in the $j$-th step.

In order to establish the locality of the fourth root of the free staggered
operator we begin with the taste representation~(\ref{Dfree}).
The lattice spacing is $2a$. Applying $n$ blocking transformations,
we obtain a coarse-lattice spacing $a_c= 2^{n+1}a$.
The blocking parameter is fixed to be $\a^{(j)}=\a$ where $\a=O(1/a_c)$,
for all steps. We will take the limit $n\to\infty$ while holding $a_c$ fixed.
Thus the original lattice spacing vanishes like $2^{-(n+1)}$.
For any finite $n$, the generalization of Eq.~(\ref{RG}) gives
\begin{equation}
  {\rm det}(D_0) = {\rm det}(G_n^{-1})\, {\rm det}(D_n)\,.
\label{fctr}
\end{equation}
It can be shown \cite{cmp} that $D_n$, $G_n$, and $G^{-1}_n$,
are all local operators on the coarse lattice,
\ie their kernels decay exponentially with an $O(1/a_c)$ decay rate,
uniformly in $n$. The limit $n\to\infty$ exists, and we find
${\rm det}(D_0)= {\rm det}(G_\infty^{-1}) \, {\rm det}(D_\infty)$,
in obvious notation.

In the massless case the blocked propagators
have the general form
\begin{equation}
  D_n^{-1}(p) = \a_n^{-1}
  -\sum_\m \Big( i[\g_\m \otimes I] \ca_\m^n(p)
  + [\g_5 \otimes \t_5 \t_\m] \cb_\m^n(p) \Big) \,,
\label{slv}
\end{equation}
where $\a_n = 15\a/(16(1-2^{-4n}))$.
For small $pa_c$, one has $\ca_\m^n(p) = (p_\m/p^2)(1+O(pa_c))$.
This corresponds to a wave-function renormalization equal to unity,
as it must be for a free theory. In units of $a_c$,
the taste-violating amplitudes $\cb_\m^n(p)$ scale like $a/a_c \propto 2^{-n}$,
uniformly in $p$.
The taste breaking originates from an irrelevant operator
(the last term in Eq.~(\ref{Dfree})), and this is indeed the scaling anticipated
for such an operator. Hence $\cb_\m^\infty(p)=0$, and $D_\infty$ factorizes as
$D_\infty = \Drg \otimes I$.
The fourth root is ${\rm det}^{1/4}(D_\infty) ={\rm det}(\Drg)$.
Being a fixed-point operator (in the massless case), $\Drg$ satisfies
the Ginsparg-Wilson relation \cite{fp}.

The operator $G^{-1}_\infty$ has an $O(1/a_c)$ gap.
The Dirac operator $D_0$ has a zero at $p=0$ (only)
and, for any $n$, this zero is lifted when adding the
blocking-kernel part whose maximum is obtained at $p=0$ (see Eq.~(\ref{DGn})).
As a result, there exists a local fourth-root operator $\cm$ satisfying
${\rm det}^{1/4}(G_\infty^{-1}) ={\rm det}(\cm)$ having a similar gap,
whose kernel has an $O(1/a_c)$ decay rate too.

\sssec{3}{Interacting staggered fermions}
Writing the interacting, one-component staggered action as $\bc (\Dst+m) \c$,
the fourth-root trick amounts to the prescription
(keeping to a one-flavor theory for notational simplicity)
\begin{eqnarray}
  \svev{\c(x)\, \bc(y) \cdots}
  &=&
  Z^{-1} \int \cd U\, {\rm det}^{1/4}(\Dst+m)\; e^{-S_g}\;
  (\Dst+m)^{-1}(x,y) \cdots\,,
\label{4th}
\NXT
  Z &=& \int \cd U\, {\rm det}^{1/4}(\Dst+m)\; e^{-S_g} \,.
\label{Z4th}
\end{eqnarray}

An essential element of the RG program is the emergence of the intermediate
scale $a_c$, which defines the separation between what we consider
as short-distance and long-distance physics. In the interacting theory
we will again aim at generating through RG blocking a new lattice theory,
whose spacing $a_c$ satisfies $a \ll a_c \ll \L_{QCD}^{-1}$.
The limit $n\to\infty$ at fixed $a_c$ (where $a/a_c\to 0$) will again,
hopefully, give rise to a local lattice theory on the coarse lattice
that, at the same time, will reproduce the physical observables of the
original rooted staggered theory. We now list the main issues that arise in the
interacting theory. A detailed account will appear elsewhere \cite{prep}.

\noindent$\bullet$
{\it Gauge-field blocking and gauge invariance.}
The gauge-field blocking can be done in the usual way, see \eg ref.~\cite{ph}.
The blocking of all fields must respect gauge invariance. For example, suitable
parallel transporters \cite{FM} must be introduced into the fermion
blocking kernel.

\noindent$\bullet$
{\it Hypercubic invariance.} For blocking of $2^4$ hypercubes, it is not
possible to construct gauge-covariant blocking kernels that
will transform covariantly under hypercubic rotations as well.
A projection onto hypercubic invariant observables can nevertheless
be enforced. It is equivalent to introducing at each blocking step
an action-less local field that lives on the coarse lattice;
its (discrete) value at a given hypercube sets the relative coordinates
of the site to which we will parallel transport all the variables residing
inside that hypercube.

\noindent$\bullet$
{\it Observables.} The observables of the coarse-lattice theory
form a subset of the original observable.
The RG transformation(s) define a local mapping of the coarse-lattice
observables back into the original fine lattice. This relation
is important because, in the absence of a normal path integral,
the fourth-root prescription is defined from the outset
in terms of observables, cf.\ Eq.~(\ref{4th}) above.

\noindent$\bullet$
{\it From one-component to taste.} As already mentioned, the interacting
theory is originally defined in the one-component formalism.
We use the resemblance between Eqs.~(\ref{chitoq}) and~(\ref{Q})
to perform a first, special, blocking step. This step realizes
a covariant version of Eq.~(\ref{chitoq}), thus defining a taste-basis
version of the interacting theory \cite{rg,FM}.
Since the observables of this interacting theory form a subset of
the original observables, in effect all the symmetries of the one-component
theory remain intact. The disastrous mass terms, which are generated
if one gauges the taste representation in a simple-minded way
\cite{MW}, are avoided.

\noindent$\bullet$
{\it ``Postponing'' multi-fermion interactions.} Equation~(\ref{fctr})
plays the key role of providing the short-distance -- long-distance
separation at the level of the fermion determinant. A technical
difficulty is that, as soon as we RG block the gauge fields even once,
multi-fermion interactions are generated, and any fermionic path
integral is no longer a determinant. The simple solution, valid for
any finite number of blocking steps, is to perform all the fermion
blocking steps ahead of all the gauge-field blocking steps.
In this context one can still make use of Eq.~(\ref{fctr}).

\noindent$\bullet$
{\it Scaling.} It will be unrealistic to expect rigorous
proofs in the interacting case. With all the above elements,
what one achieves is a framework for constructing an RG blocked
version of the interacting, rooted staggered theory.
The familiar scaling behavior, derived using perturbation theory,
should then apply for large-enough number of blocking steps $n$,
(equivalently, small-enough $a$).
This should imply in particular that all the taste-breaking effects,
that always originate from irrelevant operators, will again be damped
like $a/a_c \propto 2^{-n}$. While not a proof, this suggests
that, for $n\to\infty$, the RG blocked lattice theory emerging from
the rooted staggered theory is local.

\sssec{4}{Concluding remarks}
An advantage of the RG program is that, since it is formulated
in a completely non-perturbative language, it is amenable to numerical tests.
Numerical evidence for the (approximate) four-fold degeneracy of the low-lying
staggered eigenmodes may be found in ref.~\cite{4fold}.
For first numerical tests of the scaling properties of the RG blocked
fermion propagator in the interacting rooted theory, see ref.~\cite{FM}.

{\it Acknowledgements.}
I thank Claude Bernard, Carleton Detar, Maarten Golterman and Francesca
Maresca for numerous valuable discussions.
This work is supported by the Israel Science Foundation under grant
no. 222/02-1.


\end{document}